\documentclass[preprint]{aastex}

\usepackage{amsmath}
\usepackage[normalem]{ulem}
\usepackage[table]{xcolor}  
\usepackage{hyperref}

\begin{document}

\title{Evidence for a dual population of neutron star mergers from short Gamma-Ray Burst observations}
\shorttitle{Dual population of short GRBs}
\author{K. Siellez}
\affil{Center for Relativistic Astrophysics and School of Physics, Georgia Institute of
Technology, Atlanta, GA 30332, USA}
\affil{ARTEMIS UMR 7250, CNRS
Universit\'e Nice Sophia-Antipolis, \\ Observatoire de la C\^{o}te d'Azur, CS 34229 F-06304 NICE, France }
\email{karelle.siellez@ligo.org}
\and
\author{M. Bo\" er}
\affil{ARTEMIS UMR 7250, CNRS
University of Nice Sophia-Antipolis, \\ Observatoire de la C\^{o}te d'Azur, CS 34229 F-06304 NICE, France }
\email{Michel.Boer@unice.fr}
\and
\author{B. Gendre}
\affil{University of the Virgin Islands, College of Science, 2 John Brewer's Bay, \\
00802 St Thomas, VI, USA} 
\affil{Etelman Observatory, 00802 St Thomas, VI, USA} 
\affil{ARTEMIS UMR 7250, CNRS
University of Nice Sophia-Antipolis, \\ Observatoire de la C\^{o}te d'Azur, CS 34229 F-06304 NICE, France }
\email{Bruce.Gendre@uvi.edu}
\author{T. Regimbau}
\affil{ARTEMIS UMR 7250, CNRS
University of Nice Sophia-Antipolis, \\ Observatoire de la C\^{o}te d'Azur, CS 34229 F-06304 NICE, France }
\email{Tania.Regimbau@oca.eu}

\date{\today}% It is always \today, today,

% % % % % % % % % % % % % % % % % % % % % % % % % % % % % % % % % % % % % % % % % % % % % % % % % % %
% % % % % % % % % % % % % % % % % % % % % % % % % % % % % % % % % % % % % % % % % % % % % % % % % % %
% % % % % % % % % % % % % % % % % %           ABSTRACT           % % % % % % % % % % % % % % % % % % %
% % % % % % % % % % % % % % % % % % % % % % % % % % % % % % % % % % % % % % % % % % % % % % % % % % %
% % % % % % % % % % % % % % % % % % % % % % % % % % % % % % % % % % % % % % % % % % % % % % % % % % %

\begin{abstract}
Short duration Gamma-Ray Bursts are thought to originate from the coalescence of neutron stars in binary systems. They are detected as a brief ($<$ 2s), intense flash of gamma-ray radiation followed by a weaker, rapidly decreasing afterglow. They are expected to be detected by Advanced LIGO and Virgo when their sensitivity will be low enough. In a recent study \citep{reg15} we identified a population of short Gamma-Ray Bursts that are intrinsically faint and nearby. Here we provide evidence in favor of the existence of this new population that can hardly be reproduced with a model of field neutron star binary coalescences. We propose that these systems may be produced dynamically in globular clusters, and may result from the merger of a black hole and a neutron star. The advanced LIGO and Virgo observation of a high rate of NSBH mergers compatible with the dynamical formation in globular clusters would be a confirmation of this hypothesis and would enable for the derivation of the mass function of black holes inside globular clusters, as well as the luminosity function of faint short GRBs.
\end{abstract}


\maketitle

% % % % % % % % % % % % % % % % % % % % % % % % % % % % % % % % % % % % % % % % % % % % % % % % % % %
% % % % % % % % % % % % % % % % % % % % % % % % % % % % % % % % % % % % % % % % % % % % % % % % % % %
% % % % % % % % % % % % % % % % % %     I   -    INTRODUCTION      % % % % % % % % % % % % % % % % %
% % % % % % % % % % % % % % % % % % % % % % % % % % % % % % % % % % % % % % % % % % % % % % % % % % %
% % % % % % % % % % % % % % % % % % % % % % % % % % % % % % % % % % % % % % % % % % % % % % % % % % %

\section{Introduction}\label{Introduction}
Binary compact objects are among the best laboratories to test General Relativity \citep{ein16} and alternative theories, as well as the physics in strong field regime.
Compact binary systems include neutron star binaries (hereafter NSNS), systems with a neutron star and a black hole (NSBH), and of two black holes (BHBH). The existence of the later has been dramatically confirmed through the Advanced LIGO detection of GW 150914 \citep{GW150914}.

NSNS systems are thought to be the most plausible progenitor of short Gamma-Ray Bursts \citep[sGRBs - ][]{pacz86,Eichler89, nar92,ber14}. Once the binary system has radiated away all its angular momentum through the emission of gravitational waves, it undergoes a merging process that forms an accretion disk and an ultra-relativistic jet of matter \citep{pacz86, Eichler89}. It is not clear whether this results directly into the formation of a black hole or if an intermediate object, called a magnetar, is formed \citep{DAI2006,Metzger2008}. During the coalescence gravitational and electromagnetic waves are emitted in large amount; internal shocks within the jet produce the prompt GRB that lasts less than a second in the rest frame (see below) and whose EM emission peaks in the hard X-ray - gamma-ray range; the interaction of the jet with the surrounding medium results in the afterglow emission decreasing approximately by an order of magnitude by decade of time \citep{Mesza2012, gue12}.

The formation of compact binary systems follows a complex path as most of the mass of the stellar progenitors is released during the collapse of the individual compact objects, and the loss of about half of the binary mass leads to dynamical disruption \citep{ver95}. In order to keep the system bound during the whole process, a complicated exchange of mass occurs between the two components of the binary, possibly resulting in a common envelop stage \citep{tau03}. This results in some constraints on the final properties of the binary such as the separation or the eccentricity of the orbit \citep[][]{ver95}. The second solution is to form the binary once the two compact objects are formed, through dynamical interactions \citep{ver87, ver03}. The probability for such an interaction to occur is extremely low in the disk of normal galaxies, but large within dense environments such as globular clusters \citep{rodri16a,cha16}.

Both paths seem unlikely, but have proven true by various observations: e.g. the observation of binary pulsars \citep{hul75}, or the over-representation of low mass X-ray binaries within globular clusters \citep{heg75}. However, it is not clear what phenomenon is the most common. Usually, the direct formation mechanism is considered as such, and the dynamical process is expected marginal. As one of the challenges facing astrophysics is to detect these events both from their gravitational (hereafter refered as GW for Gravitational Waves) and electromagnetic (EM) radiation, it is of paramount importance to understand when and how each of these phenomena is at work.

With the Advanced LIGO \citep{aas15} already in operations, soon joined by Advanced Virgo \citep{ace15} (hereafter both experiment are designed as ALV, for Advanced LIGO and Virgo) the study of compact binary systems will know a new enlightenment age. Already, the detection of the coalescence of a binary system of black holes of initial masses $36_{-4}^{+5}$ M$_{\odot}$ and $29_{-4}^{+4}$ M$_{\odot}$ at a distance of $410^{+160}_{-180}$ Mpc \citep{GW150914} allows to draw astrophysical implications \citep{GW150914AST} as well as to better constrain the rates of such coalescences \citep{GW150914RATES}.

When at full sensitivity, ALV will be able to detect the merger of two neutron stars (NSNS) up to $\approx$ 480 Mpc if it is face-on and in an optimal position with respect to the detector (horizon) or $\approx$ 200 Mpc for a source averaged over sky position and source orientation (range); the merger of a neutron star and a black hole (BH of 10 M$_\odot$) can be detected up to $\approx$ 900 Mpc (horizon)  or $\approx$ 400 Mpc  (range) \citep{Abadie10}.  The ALV collaboration has estimated the rate of detections of NSNS coalescences to be between 0.2 and 200 events/year, using the three detectors \citep{abb16}. The estimates for the source density are quite diverse in the literature and respectively to 92 -- 1154 Gpc$^{-3}$ yr$^{-1}$ \citep{sie14}, 8 -- 1800 Gpc$^{-3}$ yr$^{-1}$ \citep{cow12} and 500 -- 1500 Gpc$^{-3}$ yr$^{-1}$ \citep{pet13}  have been proposed from observational constrains, while \cite{gue05} proposes the interval 8 -- 30 Gpc$^{-3}$ yr$^{-1}$ on theoretical grounds. In a recent work \citep{reg15} we estimated the number of coincident detections between GW antenna, both ALV and the proposed Einstein Telescope (ET) \citep{pun10}, and EM high-energy detectors (\textit{Swift}, Fermi-GBM and the future SVOM spacecraft) from realistic population synthesis of field binary mergers. We used various star formation rates (SFRs) and delay times between the formation of the system and its coalescence, including both GWs and sGRBs selection effects. We estimated the rate of coincident EM/GWs detections to be between 0.001 and 0.46 yr$^{-1}$, leading to a detection rate for ALV within its range between 2.5 and 3.0 yr$^{-1}$ for NSNS binary coalescence. All these numbers assume the dynamical formation of binaries to be marginal.

In this paper we present new estimates of these rates using new simulations that we performed with a new, carefully selected, sample of short GRBs. We discuss the possibility that at least part of the binary coalescences originate from system that experiences dynamic formation in dense environments.

This paper is organized as follows: in section \ref{sample} we present the method used to select the sample of sGRBs we used; in section \ref{simulations} we present the simulations and the various parameters we have used to perform them;  we discuss the results in section \ref{discussion} as well as the comparison with the observational data from our sample. Finally in section \ref{conclusion} we summarize our main conclusions and findings. Through this paper we assume a standard flat cold dark matter ($\Lambda_{CDM}$) model for the Universe, with $\Omega_m$ = 0.3 and $H_0$ = 70 km s$^{-1}$ Mpc$^{-1}$.

% % % % % % % % % % % % % % % % % % % % % % % % % % % % % % % % % % % % % % % % % % % % % % % % % % %
% % % % % % % % % % % % % % % % % % % % % % % % % % % % % % % % % % % % % % % % % % % % % % % % % % %
% % % % % % % % % % % %           II   -      Sample selection      % % % % % % % % % % % % % % % % % %
% % % % % % % % % % % % % % % % % % % % % % % % % % % % % % % % % % % % % % % % % % % % % % % % % % %
% % % % % % % % % % % % % % % % % % % % % % % % % % % % % % % % % % % % % % % % % % % % % % % % % % %

\section{Sample selection}\label{sample}

\begin{figure}[!ht]
%\plotone{f1.eps}
\includegraphics [width=\columnwidth]{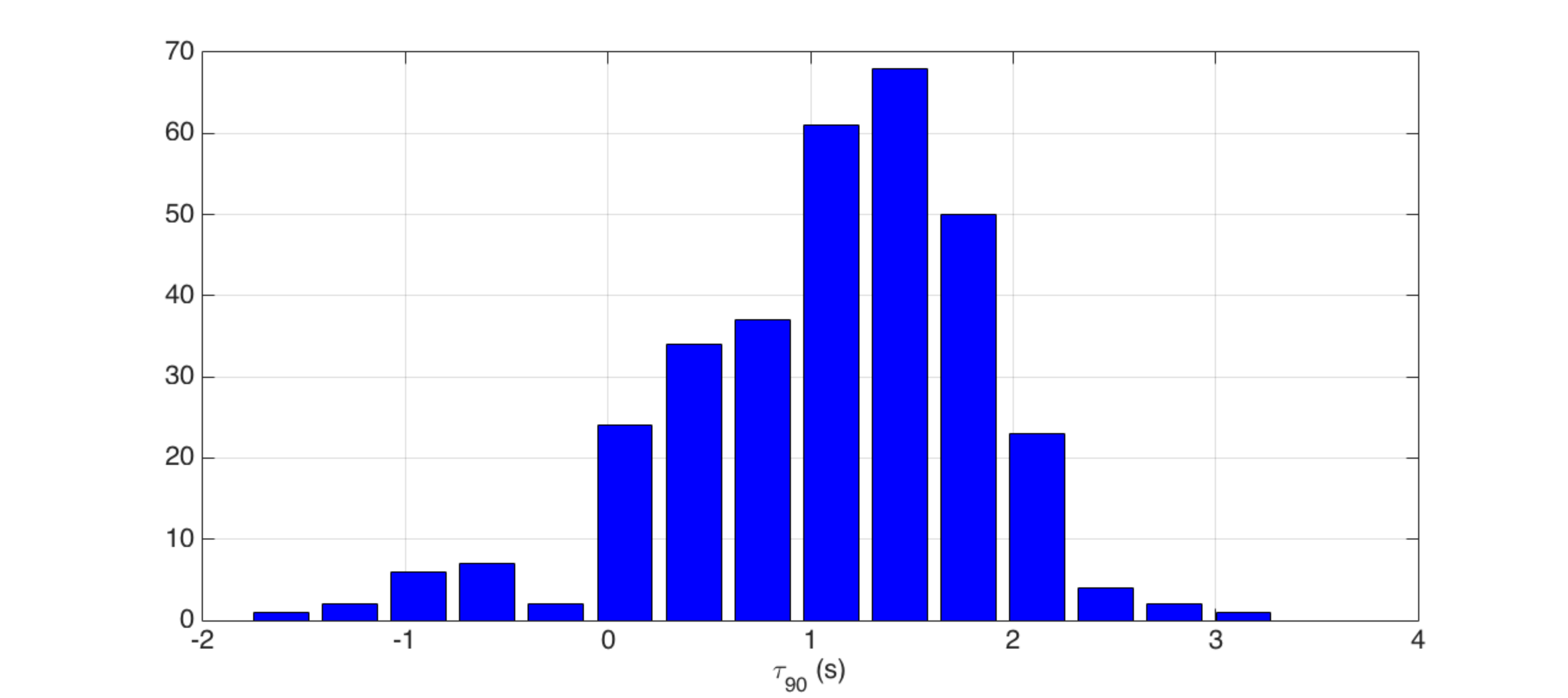}
\caption{Histogram of the $\tau_{90}$ duration for 301 GRBs from \textit{Swift}. $\tau_{90}$ is the $T_{90}$ duration computed in the rest frame.
\label{fig1}}
\end{figure}

The total number of GRBs detected by the \textit{Swift} satellite from 2004 December, 17th to 2016, May 17th is 1050 events; among them 328 have a known redshift according to the \textit{Swift} repository\footnote{\url{http://swift.gsfc.nasa.gov/archive/grb_table.html/}}, and 279 have a redshift measurement derived from the spectroscopy of the source or of the host when the association is firm. We used these redshifts to correct the $T_{90}$ (measured in the observer frame) to the $\tau_{90}$ in the rest frame. From the examination of the $\tau_{90}$ histogram (Figure \ref{fig1}) it is clear that the dividing line between short and long GRBs is 1 second {\it in the rest frame}.

The next step of the selection is based on the hardness of the energy spectrum as sGRBs are known for being harder on average than long GRBs \citep{dez92,kou93}. We used the Band model \citep{ban93}: it is a broken power law smoothly joined at a break energy called $E_b$ and defined as $E_b=(\alpha +\beta)E_0$, where $\alpha$ and $\beta$ are respectively the first and the second power law index and $E_0$ the e-folding energy \citep{kan06}. The typical value of $\alpha$ is of the order of 1.2 and $\beta$ of the order of 2.3 \citep{pre98}. Because of its limited energy band, BAT detects only the soft segment $\alpha$. The break energy is above the BAT high-energy limit. We consider a burst as hard if the measured spectral index by \textit{Swift} is lower than 2, leaving only 20 events in our sample. 

We have added 10 GRBs incorrectly flagged as "unknown redshift" in the \textit{Swift} database from a careful examination of the literature and the GCNs. 
Finally, we look at the data from other satellites and found another burst from HETE, GRB 050709 \citep{boe05} with a known redshift.

 The final sample amounts to 31 sGRBs with known redshift that verify both the temporal and spectral criteria. The data from the final sample are listed Table \ref{table1}. The first column is the burst name, followed by the experiment(s) that detected the event, then its redshift, the observed and rest frame durations, and the peak luminosity in the rest frame computed following the method described in \cite{reg15}: we used a power-law model and a Band function \citep{ban93} (when it was possible) to fit the data from \textit{Swift}, and whenever available the \textit{Fermi}-GBM and \textit{Konus} data. We finally applied the K correction to correct for the distance.

\begin{table*}[!ht]
\scriptsize
\centering
\begin{tabular}{l||ccc|cc|cc|c}
%\hline 
sGRB name &  \multicolumn{3}{c}{Instrument}  & Redshift & Ref. & \multicolumn{2}{c}{Duration [s]}    & L$_{peak}$  \\
    &      BAT & GBM	 & KONUS  &   & & Observed & Rest frame          &  [erg s$^{-1}$] \\
\hline
% & & & & & & & & \\
150423A	  & X &   &  & 1.394 & S & 0.22	 & 0.09			& $4.01 \pm 0.45 .10^{50}$\\ %& 0.84, PL (0.24)  
150120A   & X & X &   & 0.460	 & S & 1.20  	 & 0.82  		   & $9.01 \pm 1.00 .10^{49}$ \\ %& 1.81, PL (0.18)  
141212A	 & X & & & 0.596  & S & 0.30  	 & 0.19 	  	   & $1.02 \pm 0.17 .10^{50}$ \\ %& 1.61, PL (0.23) 
140903A	 & X & & & 0.351 & S & 0.30	 & 0.22 		   & $6.09 \pm 0.49 .10^{49}$ \\ % & 1.99, PL (0.12)   
131004A	 & X & X & & 0.088 & 1 & 1.54	 & 0.90 		     &$4.24 \pm 0.25 .10^{50}$\\ % & 1.81, PL (0.11)   
130603B	 & X & & X & 0.356	 & S & 0.18	 & 0.13 	  & $1.94 \pm 0.09 .10^{50}$ \\ % 	& 0.82, PL (0.07)   
120804A 	 & X & & X & $1.3^{+0.3}_{-0.2}$ & 2 & 0.80  	 & 0.35 & $4.46 \pm 0.25 .10^{51}$ \\ %& 1.34, PL (0.08)   
111117A	 & X & X & & 	$1.31_{-0.23}^{+0.46}$ & 3,4 & 0.47	 & 0.20   & $4.25 \pm 0.64 .10^{50}$ \\ %& 0.65, PL (0.22)   
101219A	 & X &  & X & 0.718  & S  & 0.60	 & 0.35 		& $4.67 \pm 0.23 .10^{50}$ \\ %& 0.63, PL (0.09)   
100724A	 & X & & & 1.288 & S & 1.40	 & 0.61 		   & $1.19 \pm 0.13 .10^{51}$ \\ %& 1.92, PL (0.21)   
100625A	 & X & X & X & 0.452 $\pm$ 0.002 & 5 	 & 0.33	 & 0.23  & $1.26 \pm 0.10 .10^{50}$ \\ %& 0.90, PL (0.10)  
100206A	 & X & X &  & 0.41 & S & 0.12	 & 0.08 	& $5.66 \pm 0.81 .10^{49}$ \\  %& 0.63, PL (0.17) 
100117A	 & X & X & & 0.915 & 6 & 0.30	 & 0.16	& $5.03 \pm 0.70 .10^{50}$ \\ %& 0.88, PL (0.22)  
090927	 & X & X & & 1.37 & S  & 2.20	 & 0.93     & $8.88 \pm 0.89 .10^{50}$ \\ %& 1.80, PL (0.20) 
090515	 & X & &  & 0.403 & 7 & 0.036 	 & 0.026	& $2.03 \pm 0.32 .10^{50}$ \\ %& 0.05, CPL (1.36) 
090510	 & X & X & X & 0.903 & S & 0.30 & 0.16 & $1.43 \pm 0.17 .10^{51}$ \\ %& 0.98, PL (0.20) 
090426	 & X & & & 2.609 & S & 1.20	 & 0.33 & $6.93 \pm 0.87.10^{51}$  \\  %& 1.93, PL (0.22)   	
\rowcolor{gray}090417A	 & X & &  & 0.088 & 1 & 0.072 	 & 0.066	 	 & $4.18 \pm 1.16 .10^{48}$ \\ %& 0.65, CPL (2.11) 
\rowcolor{gray}080905A	 & X & X  & & $0.1218 \pm 0.0003$ & 8   & 1.00 & 0.89 & $4.92 \pm 0.76 .10^{48}$ \\ %& 0.85, PL (0.24) 
\rowcolor{gray}070923	 & X & & & 0.076 & 9 & 0.05	 & 0.046 & $2.89 \pm 0.48 .10^{48}$ \\ %& 1.02, PL (0.29)   
070729	 & X & & X & 0.27 (n/a)        	& 10 & 0.90 & 0.50	  & $ 1.56 \pm 0.32 .10^{50}$ \\ %& 0.8 $\pm$ 0.1 
070724A	 & X & & & 0.457  & S & 0.40	 & 0.27   & $4.59 \pm 0.92 .10^{49}$ \\ %& 1.81, PL (0.33)
070429B	 & X & & & 0.904 & S & 0.47	 & 0.25   &  $4.00 \pm 0.55 .10^{50}$ \\ %& 1.72, PL (0.23) 
061217	  & X & & X & 0.827 & S & 0.21	 & 0.11  & $2.48 \pm 0.40 .10^{50}$ \\ %& 0.86, PL (0.30) 
\rowcolor{gray}061201	 & X & & & 0.111 & S & 0.76	 & 0.68   & $1.18 \pm 0.10 .10^{49}$ \\ %	& 0.81, PL (0.15) 
060801	 & X & & & 1.1304	& 11 & 0.49	 & 0.23 	  & $4.53 \pm 0.57 .10^{50}$ \\ %& 0.47, PL (0.24) 
\rowcolor{gray}060502B	 & X & & & 0.287 & S   & 0.13  	 & 0.10 	  	 & $1.05 \pm 0.204 .10^{49}$ \\ %& 0.98, PL (0.19) 
051221A	 & X & & X & 0.547 &	S & 1.40	 & 0.90  & $8.65 \pm 0.29 .10^{50}$ \\  %& 1.39, PL (0.06) 
050813	 & X & & & 1.8 & S  & 0.45	 & 0.16 		 & $8.13 \pm 1.99 .10^{50}$ \\ %& 1.28, PL (0.37)  
050709	 & & HETE2& & 	 0.16 &   12 & 0.07	 & 0.06 	 & $5.2 \pm 1.4 .10^{50}$ \\
\rowcolor{gray}050509B	 & X & & & 0.225 &	S  & 0.073 	& 0.060 	& $3.00 \pm 1.07 .10^{48}$ \\ %& 1.57, PL (0.38)  
\end{tabular}
\caption{The sample of sGRBs with the name, the detector(s) that has observed it, the redshift, the observed $T_{90}$ and rest frame $\tau_{90}$ durations, the peak luminosity computed in the rest frame. The 6 low-z faint sGRBs are highlighted in gray (see text). The references for the redshift are given by: S for the \textit{Swift} data center archive : \url{http://swift.gsfc.nasa.gov/archive/grb_table.html/}; [1] : \protect\cite{obr09,fox09,blo09},  [2] : \protect\cite{ber13}, [3] : \protect\cite{sak13}, [4] : \protect\cite{mar12}, [5] : \protect\cite{fon13}, [6] : \protect\cite{fon11}, [7] : \protect\cite{ber10}, [8] : \protect\cite{row10}, [9] : \protect\cite{fox07}, [10] : \protect\cite{lei10}, [11] : \protect\cite{ber07}, [12] : \protect\cite{fox05}.}
\label{table1}
\end{table*}

\begin{figure}[!ht]
\includegraphics [width=\columnwidth]{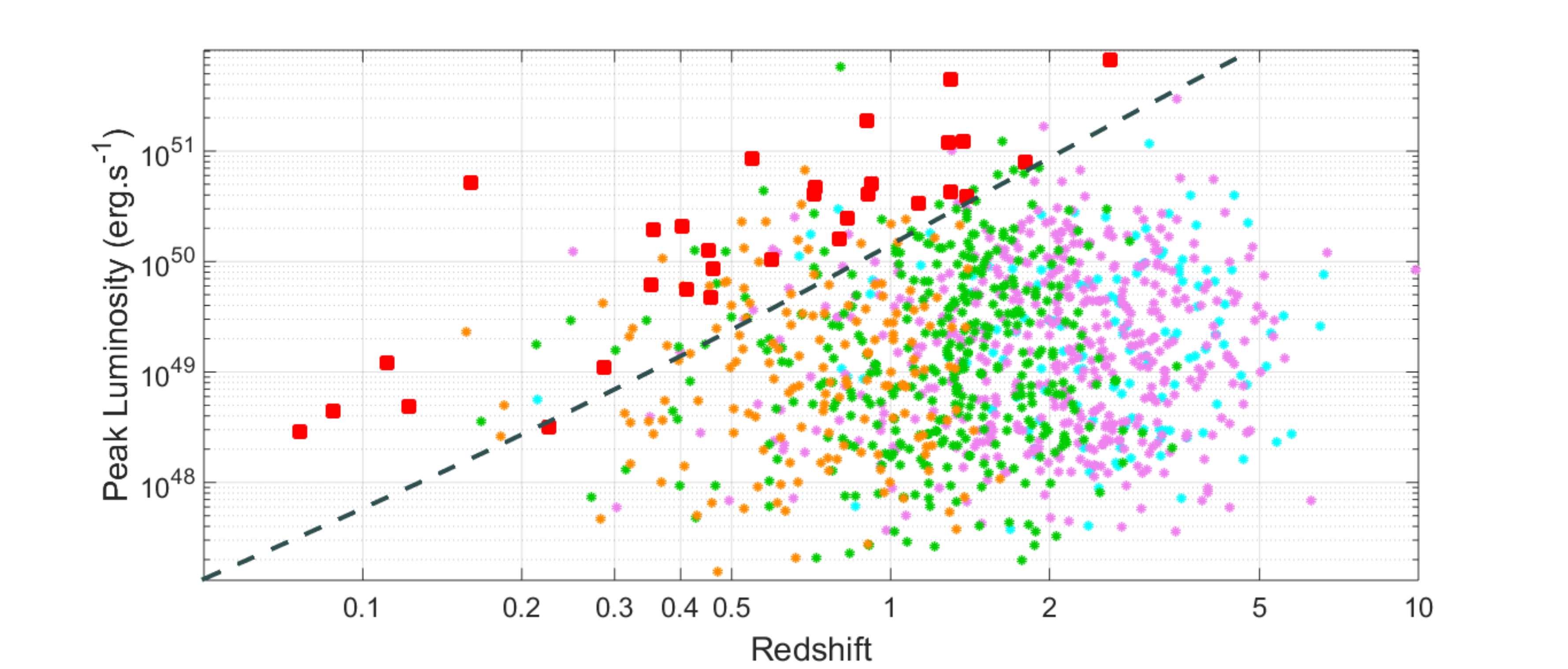}
\caption{The peak luminosity of sGRBs as a function of the redshift for the SFR model of \cite{hop06} for the 31 events from our sample (red squares) and the simulated population (dots). The colors indicate the different delays used: 20 Myr is in cyan, 100 Myr in purple, 1 Gyr in green and 3 Gyr in orange.
\label{fig5}}
\end{figure}

% % % % % % % % % % % % % % % % % % % % % % % % % % % % % % % % % % % % % % % % % % % % % % % % % % %
% % % % % % % % % % % % % % % % % % % % % % % % % % % % % % % % % % % % % % % % % % % % % % % % % % %
% % % % % % % % % % % %           II   -      Monte-Carlo simulations      % % % % % % % % % % % % % % % % % %
% % % % % % % % % % % % % % % % % % % % % % % % % % % % % % % % % % % % % % % % % % % % % % % % % % %
% % % % % % % % % % % % % % % % % % % % % % % % % % % % % % % % % % % % % % % % % % % % % % % % % % %

\section{Monte-Carlo simulations}\label{simulations}

The method used to perform the simulations and to compare them with the actual data is described in details in \cite{reg14}. 
 The purpose is to simulate a population of binary systems and to follow their evolution until they merge, resulting possibly in the detection of a sGRB or/and of a GW event by ALV.
We assume that the progenitor of the sGRBs is due to field binaries: two massive stars linked in a binary system that survived two core collapses to form a compact pair system. The two main candidates for the production of sGRBs are either NSNS or NSBH systems. We fix the masses to be 1.4 M$_{\odot}$ for neutron stars and 10 M$_{\odot}$ for black holes. Both progenitors are present in the simulations.
For each kind of progenitor, we extract the parameters of the distribution from our sample, following our previous procedure \citep{gue05,ghi09}. 
The intrinsic peak flux  (in erg s$^{-1}$) is drawn from a standard broken power law distribution:

\begin{equation}
\Phi(L_p) \propto
\left\{
\begin{aligned}
(L_p / L_*)^{\alpha} &\,\ & \mathrm{if} & \,\   L_* / \Delta_1< L_p < L_* , \\  
(L_p / L_*)^{\beta}  &\,\ &\mathrm{if} & \,\  L_*< L_p < \Delta_2 L_* ,
\end{aligned}
\right.
\end{equation}

The value of $L_*~\approx$ 1.6 x 10$^{50}$ erg s$^{-1}$ and $\alpha$ = $-0.9$ are taken from our sample. As we cannot constrain $\beta$ we take the standard value given in \citet{ghi09}, i.e. -2.3. $\Delta_1~=~\Delta_2~=~100$ so that more than 99.99$\%$ of the luminosities of our sample is covered. We need to take into account the intrinsic duration of the burst. We neglect the possible correlations between the peak luminosity $L_p$ and the duration. We fit the rest frame duration with a log-gaussian distribution of mean $\mu_{\log T_i} =-0.6629$ and standard deviation  $\sigma_{\log T_i} = 0.4156$.

We have used 3 different star formation rates: the model of \citet{hop06} is derived from measurements of the galaxy luminosity function in the ultra-violet (UV) and far infra-red (FIR) wavelengths, and is normalized by the Super Kamiokande limit on the electron antineutrino flux from past core-collapse supernovae; a more recent model \citep{van15} is constrained by the rate of GRBs at high redshifts, resulting in a slower fall-off of the SFR for large redshifts; the last model \citep{tor07} is derived from cosmological numerical simulations. We display in Figure \ref{fig5} the results of these simulations with a minimum delay of 20 Myr. 

\begin{figure}[!ht]
\includegraphics [width=\columnwidth]{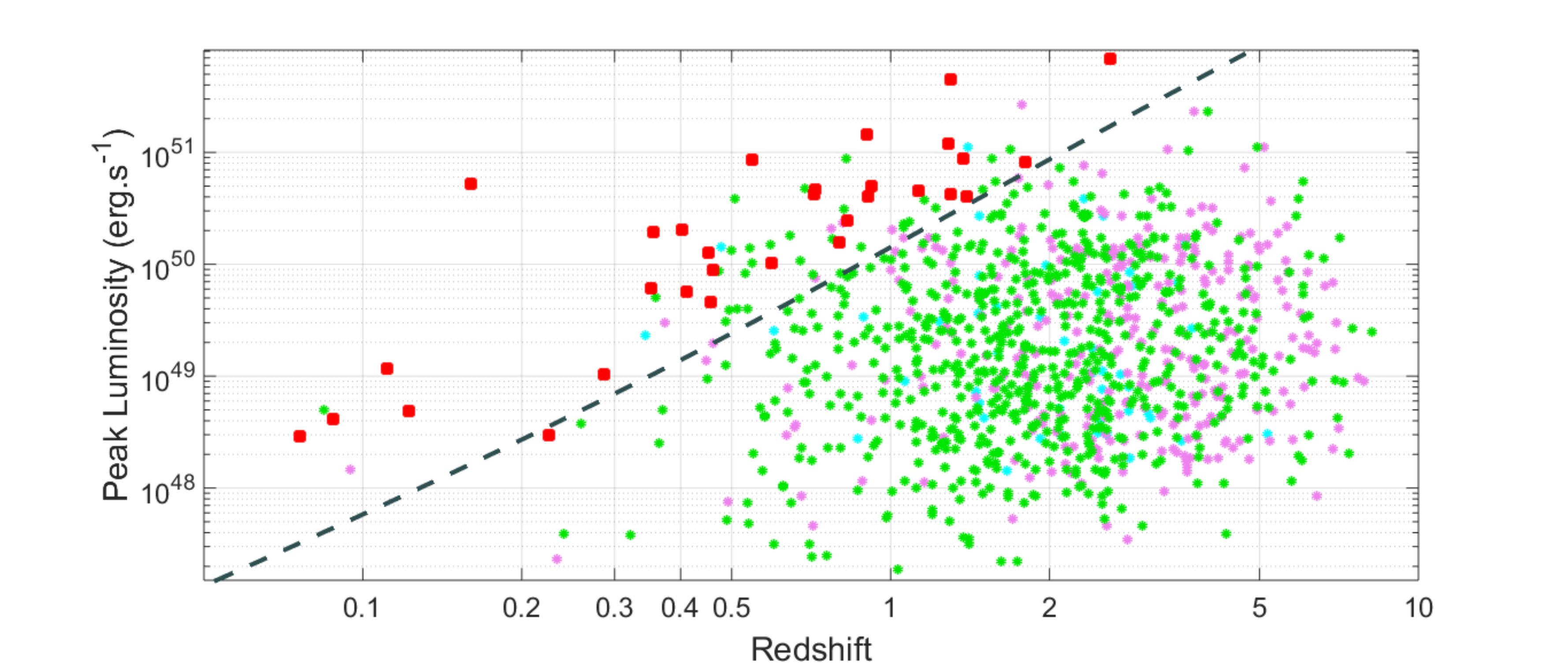}
\caption{The peak luminosity of sGRBs as a function of the redshift for the simulated population (dots) and the 31 events from our sample (red squares). The dashed line is the \textit{Swift} sensitivity limit. We have used different SFRs: cyan for \cite{hop06}, green for \cite{van15} and purple for \cite{tor07}.
\label{fig2}}
\end{figure}

We also vary the delay taken by the system to inspiral and merge. As in our previous work \citep{reg15}, P($t_d$) is the probability distribution of the delay between the formation and the coalescence. We assume a distribution in the form $P(t_d) \propto 1/t_d$  with a minimal delay of 20 Myr for the population of BNS and 100 Myr for NSBH, as suggested by the population synthesis software StarTrack \citep{dom12}. Figure \ref{fig2} shows the simulations of the sGRBs observed compared to the observational data, using the SFR from \citet{hop06} and different delays, from 20 Myr to 3 Gyr. 

\begin{figure}[!ht]
\includegraphics [width=\columnwidth]{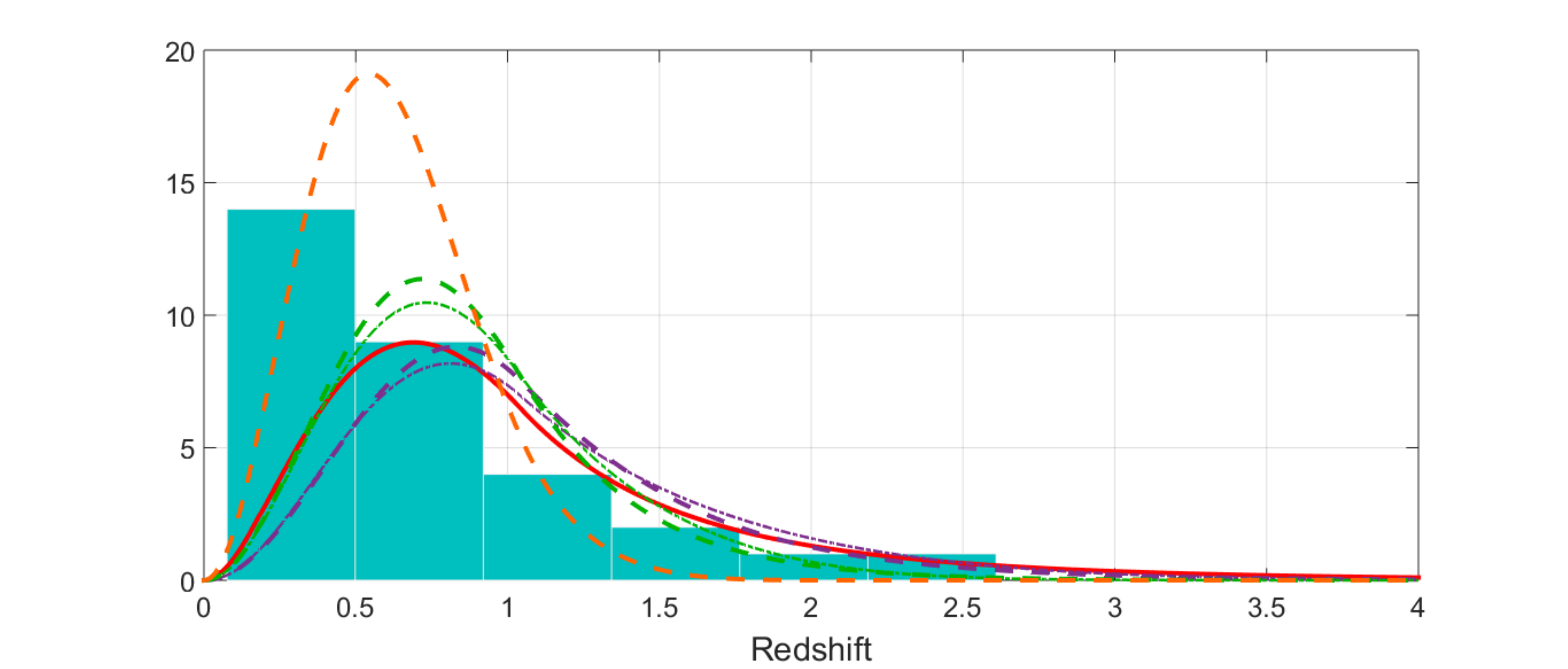}
\caption{Histogram of the redshift distribution for our sGRBs sample. The results of the simulations are plotted with the different lines for the different parameters used: red continuous line for a flat SFR; dashed-dotted line for the SFR from \cite{hop06}; dashed line for the SFR of \cite{van15}. In purple the minimum delay $t_d$ between the coalescence and the merger is 100 Myr, in green $t_d$ = 1 Gyr and in orange $t_d$ = 3 Gyr.
\label{fig3}}
\end{figure}

An examination of Figures \ref{fig5} and \ref{fig2} show that we cannot reproduce the low-redshift, low rest frame luminosity population using different combinations of SFRs and delays. Even the longest (3Gyr) fails to reproduce the nearby -- dim sources in the lower left part of the diagram. This is even more obvious in Figure \ref{fig3} where we have plotted a histogram of the sGRBs from our sample as a function of the redshift together with several simulations, varying the SFR and delay before merging. The 6 outliers are all at redshifts $<$ 0.3 and fainter on average than "regular" sGRBs. They are highlighted in gray in the Table \ref{table1}. 

\begin{figure}[!ht]
%\plotone{f1.eps}
\includegraphics [width=\columnwidth]{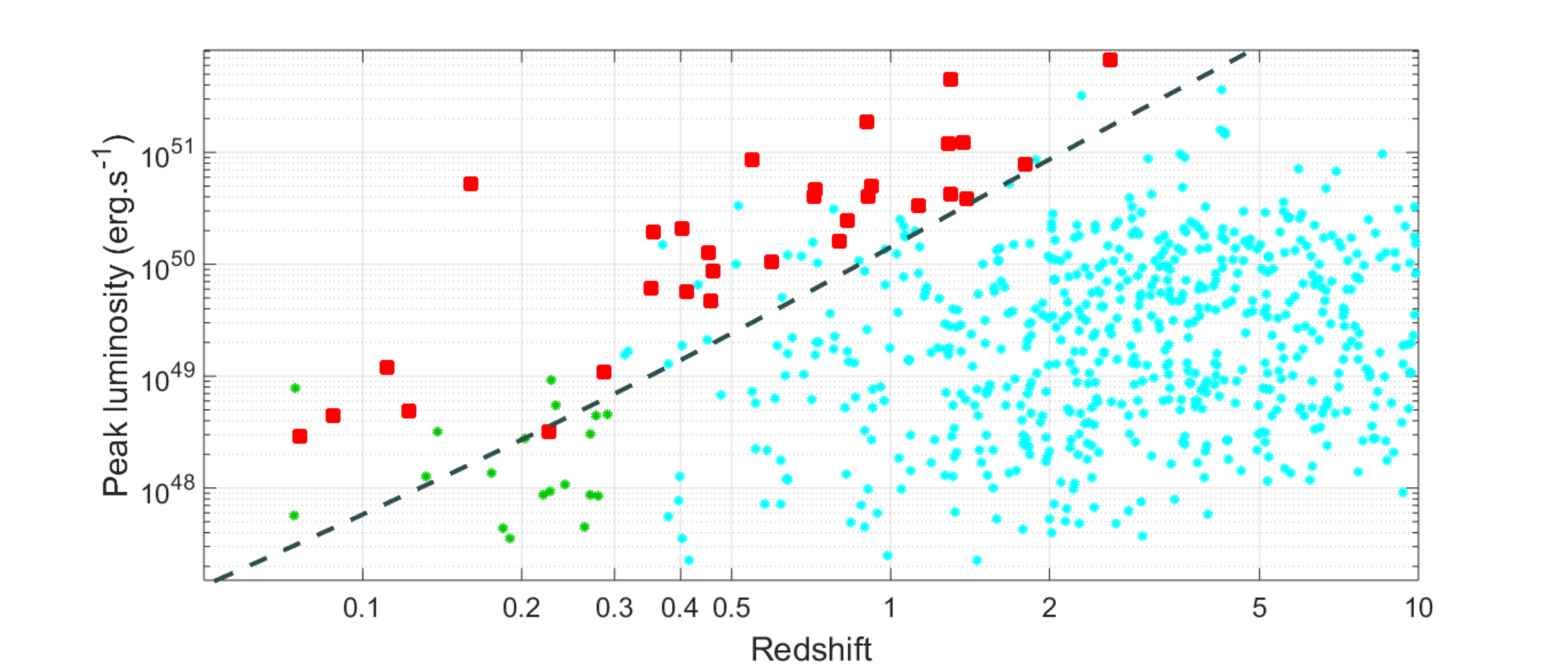}
\caption{The simulation of NSBH mergers in GCs (green dots) superposed to the field mergers (blue dots). Red dots sGRBs detected by \textit{Swift}; dashed line, the \textit{Swift} limiting sensitivity
\label{fig4}}
\end{figure}

In order to reconciliate the simulations and the data we have added a population of NSBH mergers with a luminosity 10$\%$ fainter than that of the bulk of the sGRBs population, hence with an average peak flux of $L_*~\approx$ 1.6 x 10$^{50}$ erg s$^{-1}$. At redshifts greater than 0.3 these events are undetectable for \textit{Swift}. Figure \ref{fig4} shows a combination of both populations.

We thus conclude that there is a population of 6 events that are at a lower redshift (z $<$ 0.3) and are significantly fainter than classical sGRBs.
Since these sources are faint, we see probably the tip of the iceberg. We can compute a lower limit on the rate of these faint sGRBs using the available data: 6 sources have been detected in 11.4 years of observation at a redshift smaller than 0.3 with \textit{Swift}; only 31$\%$ of the detections result in a measure of the redshift; this lead to a lower limit of the rate density for this subpopulation of 2.1 Gpc$^{-3}$ yr$^{-1}$.
If we consider a jet angle of $25^{\circ}$ and $7^{\circ}$ these numbers convert to 23 Gpc$^{-3}$ yr$^{-1}$ and 295 Gpc$^{-3}$ yr$^{-1}$  respectively. These lower limits are in agreement with the NSBH merger rate computed by \cite{van16} from r-process nucleosynthesis. 

If the origin of this population is a compact binary system, we expect several sources to be within the range of ALV during the operational life at full sensitivity. We note that during their initial version (i.e. prior to the current upgrade to the "advanced" detectors), with a 45 Mpc NSBH range \citep{aba12}, neither LIGO nor Virgo could detect them, as the volume sampled was too small. The range of Advanced LIGO during the O1 run was around 60 Mpc \citep{abb16}, still too narrow to detect these mergers. %si jamais le rate BNS NSBH est publie, rajouter la reference

% % % % % % % % % % % % % % % % % % % % % % % % % % % % % % % % % % % % % % % % % % % % % % % % % % %
% % % % % % % % % % % % % % % % % % % % % % % % % % % % % % % % % % % % % % % % % % % % % % % % % % %
% % % % % % % % % % % % % % % %%          VI   -  DISCUSSION          % % % % % % % % % % % % % % % % 
% % % % % % % % % % % % % % % % % % % % % % % % % % % % % % % % % % % % % % % % % % % % % % % % % % %
% % % % % % % % % % % % % % % % % % % % % % % % % % % % % % % % % % % % % % % % % % % % % % % % % % %

\section{Discussion}\label{discussion} 
The above figures for the density rate of the proposed NSBH mergers are lower limits. Actual figures can be much larger, prompting to search for another possible origin.

Some sGRB sources may originate from compact binary systems in globular clusters (GCs) \citep{gri06,lee10}.  They host dynamical formation of NSNS, NSBH and BHBH binaries through close-encounter of stars and bodies \citep{hut92,gen03,poo03}. In these systems, the number of binaries is proportional to the dynamical encounter rate $\propto \rho_{0}^{1.5}~r_c^2$ (where $\rho_0$ is the central density of objects and $r_c$ the core radius of the cluster). If we assume a total number N of stars within the cluster, the number of binaries is proportional to $N \sqrt{\frac{N}{r_c^5}}$, about 20N for a typical globular cluster. Moreover the dynamical interactions tend to harden the binaries \citep{heg75}, i.e. to reduce their orbital separation and to increase the mass by exchange of binary members. As a consequence, the formation of BHBH systems is favored in globular clusters \citep{bae14,rodri16a}. The rate of NSBH mergers is uncertain and depends on the BH mass function in GCs that is poorly understood \citep{cla13}. This situation may change dramatically with the recent detection of GW 150914 \citep{GW150914,GW150914AST} as we can expect more BHBH and possibly other types of compact binaries to be detected within the next few month - years, leading to more constrains on the rate of these events, the parameters of the system, and hopefully their origin (field vs. dynamic). The detection of an EM counterpart, such as the tantalizing, albeit not conclusive, detection of a weak transient by \textit{Fermi}-GBM compatible in time and localization with GW 150914 \citep{conn16}, would certainly provide a definitive answer.

Why this population has not been already observed by the various GRB detectors in space, such as the HETE-2 \citep{lam04} and \textit{Swift} \citep{geh04} missions that are able to localize high-energy transients with arcmin to arcsec accuracy ?

In the case of the merger of a NS onto a black hole, the transient accretion disk that forms can be advection dominated \citep{ich77, nar95, yua14}. This can occurs in two cases: either the accretion rate is too low for the system to generate enough viscosity, or it is too high and the energy extracted by the viscosity is stored into the disk as heat instead of being radiated \citep{esi97, abra98}. In both cases, most of the potential energy from accretion will not be radiated but swallowed into the black hole \citep{ich77, nar95}. Most of the estimates done so far indicate that only about 10$\%$ of the potential energy from accretion will be radiated \citep{yua14}. This would lead to a fainter EM emission.

To test this hypothesis we have performed new simulations. We added to the population of field NSNS mergers a population of NSBH mergers from compact binary systems produced dynamically in globular clusters, up to a distance of z $<$ 0.3 where the sources become undetectable for \textit{Swift}. The luminosity of dynamical mergers has been reduced by a factor of 10 with respect to NSNS mergers to take into account the advection-dominated accretion disk. The results of our simulations are displayed Figure \ref{fig4} and show a good agreement with our observational sample (Table \ref{table1}).

The absence of detection in our own Galaxy can be easily explained by the fact that below a redshift of 0.05 the small volume sampled and the beaming of the jet in sGRBs lead to a small probability of detection.
NSBH mergers are stronger GW emitters than NSNS ones. As the field population of NSBH mergers is poorly constrained and is not dominant, it has been believed until now that the first sources to be detected would be NSNSs. The presence of a population of dynamically produced mergers leads to revise the number of detectable sources in GW by ALV, not only for BHBH mergers \citep{rodri16b} but also for all kind of compact binary sources. However, the number of possible mergers in globular cluster and their type (NSNS, NSBH or BHBH) depends critically on the black hole mass distribution function inside globular clusters, as well as the luminosity function of these mergers.
The discovery of a new population of NSBH mergers during the operational life of ALV would provide strong constrains on the mass distribution of black holes. It is a strong prediction of our hypothesis that is easily testable with ALV.

% % % % % % % % % % % % % % % % % % % % % % % % % % % % % % % % % % % % % % % % % % % % % % % % % % %
% % % % % % % % % % % % % % % % % % % % % % % % % % % % % % % % % % % % % % % % % % % % % % % % % % %
% % % % % % % % % % % % % % % %%          VI   -  CONCLUSION          % % % % % % % % % % % % % % % % 
% % % % % % % % % % % % % % % % % % % % % % % % % % % % % % % % % % % % % % % % % % % % % % % % % % %
% % % % % % % % % % % % % % % % % % % % % % % % % % % % % % % % % % % % % % % % % % % % % % % % % % %
\section{Conclusion}\label{conclusion} 

In this paper we have presented some evidences for a new population of short GRBs that are on average closer and fainter than \textit{classical} sGRBs. These sources originate possibly from the coalescence of a binary neutron star - black hole systems that we propose are produced dynamically in globular clusters. The relative faintness of the electromagnetic events can be easily explained by advection processes that suppress a large part of the EM emission.

With the detection of GW 150914 during the early run of LIGO we can expect that more binary compact mergers will be detected during the next scientific runs that will take place soon. The GW signal from a NSBH coalescence is strong enough to be detected easily by Advanced LIGO and Virgo, providing more data on the astrophysical parameters of these sources and their mass distribution. The detection of a sGRB firmly associated with a NSBH coalescence detected by ALV would provide a strong confirmation of our hypothesis.

%************************************************%

\section*{Acknowledgments}
We thank M. Branchesi and N. Gherels for their helpful comments. This work is under the auspice of the FIGARONet collaborative network, supported by the Agence Nationale de la Recherche, program ANR-14-CE33. KS has been supported by a PhD grant from the Ecole Doctorale SF2A of the University of Nice Sophia-Antipolis, and is currently partially supported by NSF PHY-1505524 and GATECH. BG acknowledges financial support of the NASA through the NASA Award NNX13AD28A and the NASA Award NNX15AP95A. This research has made use of data, software and/or web tools obtained from the High Energy Astrophysics Science Archive Research Centre (HEASARC), a service of the Astrophysics Science Division at NASA/GSFC and of the Smithsonian Astrophysical Observatory's High Energy Astrophysics Division. 

{\it Facilities}
\facility{HETE-2}
\facility{Swift}
\facility{LIGO}

%************************************************%

%\bibliography{biblio}

%\label{lastpage}

\newpage

\end{document}